\documentclass[11pt]{article}

\usepackage{epsfig}
\usepackage{graphicx}
\usepackage{epstopdf}
\usepackage{amsfonts}
\usepackage{amssymb}
\usepackage{amsthm}
\usepackage{amsmath}
\usepackage{multirow}

\usepackage{color}

\def\beq{\begin{equation}}
\def\eeq{\end{equation}}
\def\bea{\begin{eqnarray}}
\def\eea{\end{eqnarray}}
\newcommand{\beqs}{\begin{subequations}}
\newcommand{\eeqs}{\end{subequations}}

\newcommand{\cref}[1]{Ref.~\cite{#1}}

\newcommand{\vev}[1]{\left<#1\right>}

\newcommand{\hh}{{\ensuremath{I{\kern-2.6pt h}}}}
\newcommand{\bhh}{{\ensuremath{\bar{I{\kern-2.6pt h}}}}}

\begin{document}

	\begin{titlepage}
	
	\vspace*{-15mm}
	\begin{flushright}
	{UT-STPD-19/01}\\
	\end{flushright}
	\vspace*{0.7cm}

	\begin{center}
		{\Large {\bf Monopoles, Strings, and Necklaces in $SO(10)$ and $E_6$
		}}
		\\[12mm]
		George Lazarides~$^{a}$~\footnote{E-mail: \texttt{lazaride@eng.auth.gr}},
	Qaisar Shafi~$^{b}$~\footnote{E-mail: \texttt{shafi@bartol.udel.edu}}
	\end{center}
	\vspace*{0.50cm}
	\centerline{$^{a}$ \it
		School of Electrical and
		Computer Engineering, Faculty of Engineering,
	}
	\centerline{\it
		 Aristotle University
		of Thessaloniki, Thessaloniki 54124, Greece}
	\vspace*{0.2cm}
	\centerline{$^{b}$ \it
		Bartol Research Institute, Department of Physics and 
		Astronomy,}
	\centerline{\it
		 University of Delaware, Newark, DE 19716, USA}
	\vspace*{1.20cm}
	\begin{abstract}
		
		We employ a variety of symmetry breaking patterns in 
		$SO(10)$ and $E_6$ Grand Unified Theories to demonstrate 
		the appearance of topological defects including 
		magnetic monopoles, strings, and necklaces. We show that 
		independent of the symmetry breaking pattern, a 
		topologically stable superheavy monopole carrying a 
		single unit of Dirac charge as well as color magnetic 
		charge is always present. Lighter intermediate mass 
		topologically stable monopoles carrying two or three 
		quanta of Dirac charge can appear in $SO(10)$ and $E_6$ 
		models respectively. These lighter monopoles as well as 
		topologically stable intermediate scale strings can 
		survive an inflationary epoch. We also show the appearance 
		of a novel necklace configuration in $SO(10)$ broken to 
		the Standard Model via $SU(4)_c\times SU(2)_L\times 
		SU(2)_R$. It consists of $SU(4)_c$ and $SU(2)_R$ monopoles 
		connected by flux tubes. Necklaces consisting of monopoles 
		and antimonopoles joined together by flux tubes are also 
		identified. 
		Even in the absence of topologically stable strings, a 
		monopole-string system can temporarily appear. This system 
		decays by emitting gravity waves
		and we provide an example in which the spectrum of these 
		waves is strongly peaked around $10^{-4}~{\rm Hz}$ with
		$\Omega_{\rm gw}h^2\simeq 10^{-12}$. This spectrum should be 
		within the detection capability of LISA.
		
	\end{abstract}
\end{titlepage}

\section{Introduction}

Grand Unified Theories (GUTs) such as $SU(4)_c\times SU(2)_L
\times SU(2)_R$ (422, for short) \cite{pati}, $SU(5)$ 
\cite{glashow}, $SO(10)$ \cite{minkowski}, and $E_6$ 
\cite{ramond} predict the existence of topologically stable 
magnetic monopoles \cite{monopole}. The mass and the magnetic 
charge carried by the monopoles depends on the underlying GUT 
and its symmetry breaking pattern. For instance, in breaking 
$SU(5)$ to the Standard Model (SM) gauge group, the lightest 
monopole carries one unit of Dirac magnetic charge (and also 
color magnetic charge) \cite{daniel}, and it weighs about ten 
times the GUT scale $M_{\rm GUT}$. In contrast, a 422 breaking 
to the SM yields a stable monopole with two units of 
Dirac charge \cite{magg}, and its mass depends on the scale of 
the 422 breaking which can be lower, even significantly so, 
than the standard GUT scale $M_{\rm GUT}\sim 10^{16}~{\rm GeV}$. 
Another interesting example of GUT scale and lighter monopoles 
comes from $E_6$ breaking via the trinification group $SU(3)_c
\times SU(3)_L\times SU(3)_R$ (333, for short). This breaking 
produces a GUT scale $Z_3$ monopole that carries one unit of 
Dirac magnetic charge \cite{ann}, as we shall verify later. 
The subsequent breaking of 333 to the SM gauge group 
yields a stable intermediate mass monopole which carries three 
quanta of Dirac magnetic charge \cite{ann}.

The presence of topologically stable strings in these models 
depends on the Higgs fields that are employed to implement the 
symmetry breaking. A prime example is the appearance of $Z_2$ 
strings if $SO(10)$ is broken to the SM using only 
tensor representations \cite{z2string}. The gauge $Z_2$ 
symmetry in this case happens to be subgroup of the $Z_4$ 
center of $SO(10)$. In supersymmetric $SO(10)$ this $Z_2$ is 
precisely equivalent to matter parity which, among other 
things, provides a stable cold dark matter candidate, namely 
the lightest sparticle. 

Composite topological defects can also appear in many GUTs and 
some well-known examples include monopole-antimonopole pairs 
connected by a string (dumbbells) \cite{dumbbells}, walls 
bounded by strings \cite{kibble}, and necklaces with monopoles 
acting as beads kept together on a string \cite{vachaspati}.
Consider, for 
instance, the breaking of $SO(10)$ to 422 with a {\bf 54}-plet 
of Higgs. This leaves unbroken a discrete symmetry, called 
$C$-parity, which interchanges the left and right components 
of any representation, accompanied by charge conjugation 
\cite{kibble}. Under $C$ the electric charge operator 
$Q\to -Q$ \cite{kibble,membrane}. This breaking of $SO(10)$ 
to 422 yields $Z_2$ strings. However, the subsequent breaking 
of the 422 symmetry to the SM group necessarily 
breaks this $C$-parity, and the strings form boundaries of 
domain walls \cite{kibble}. Such walls can be tolerated in 
realistic scenarios provided they are unstable and disappear 
before their energy density becomes the dominant component in 
the universe. Another well known option, if available, is to 
inflate away the domain walls. It is interesting to note that 
observation of walls bounded by strings in $^3$He has been 
reported recently in Ref.~\cite{volovik}. An example of a 
necklace made up of monopoles and antimonopoles connected by 
a $Z_2$ string is provided by the symmetry breaking $SO(10)
\to SU(5)\times U(1)\to SU(5)\times Z_2$ where the last 
step is achieved by a Higgs {\bf 126}-plet of $SO(10)$. We 
will demonstrate the appearance of a new type of necklace if 
$SO(10)$ breaking occurs via 422.

Of great interest, of course, 
is the question as to whether any of these primordial 
topological defects exist in nature, having either survived 
inflation or making an appearance after the inflationary 
epoch. It has been recognized \cite{extended,senoguz} for 
some time that monopoles associated with an intermediate 
scale $M_{\rm I}$ that is comparable to $H$, the Hubble 
scale during inflation, may be present in our galaxy at an 
observable level. This can come about if the number of
$e$-foldings experienced during the intermediate scale 
phase transition is around 25-30, rather than the 50-60 
$e$-foldings experienced by the GUT scale phase transition. 
Intermediate scale cosmic strings, on the other hand, can 
appear either in the same way as the monopoles, or even 
after the end of inflation. The current bound from Cosmic 
Microwave Background Radiation measurements on the 
dimensionless string tension is given by $G\mu_{\rm s} 
\lesssim 3.2\times 10^{-7}$ \cite{planck}, where $G$ 
denotes Newton's constant and $\mu_{\rm s}$ is 
the mass per unit length of the string. Somewhat more 
stringent constraints on $G\mu_{\rm s}$ based on 
pulsar timing observations have been reported in 
Ref.~\cite{olum}. 

In this paper, we discuss 
topological defects in GUTs, with emphasis on $SO(10)$ and 
$E_6$ (see also Ref.~\cite{king1,king2} for a recent 
discussion on related topics). In Sec.~\ref{sec:422} we show 
the presence of a GUT monopole carrying one unit of Dirac 
magnetic charge in $SO(10)$ models, which is independent 
of the symmetry breaking pattern. Analogous to the $SU(5)$ 
case, this monopole carries some color magnetic charge.
We break the $SU(4)_c\times SU(2)_L\times SU(2)_R$ 
symmetry to the SM in two steps and show how an intermediate 
mass monopole carrying two units of the Dirac charge 
(Schwinger monopole) emerges from a coalescence of $SU(4)_c$ 
and $SU(2)_R$ monopoles bound together by flux tubes in a 
dumbbell configuration.  This symmetry breaking pattern of 
422 also yields a new type of necklace configuration 
consisting of alternating $SU(4)_c$ and $SU(2)_R$ monopoles 
connected by suitable flux tubes. A variety of other 
configurations is also possible including a necklace made of
monopoles and antimonopoles connected by a $Z_2$ string. 
In Sec.~\ref{sec:333} we show the 
presence of the GUT Dirac monopole also in $E_6$ models and 
discuss the $E_6$ breaking via 333, which leads to 
intermediate scale monopoles with three units of Dirac 
magnetic charge and possibly to non-superconducting stable 
strings. In Sec.~\ref{sec:psi} we analyze the $E_6$ breaking 
via $SO(10)\times U(1)_\psi$ and show how unstable strings as 
well as stable strings or necklaces can appear. 
Sec.~\ref{sec:inf} presents a quantitative discussion of how 
intermediate scale monopoles, strings, and necklaces in 
realistic models can survive primordial inflation. In addition 
we discuss how gravity waves emitted by some defects may be 
accessible with the space based observatory LISA. Our 
conclusions are summarized in Sec.~\ref{sec:conclusion}.

\section{$SO(10)$ breaking via $SU(4)_c\times SU(2)_L\times 
SU(2)_R$}
\label{sec:422}

We will first study the breaking of $SO(10)$ via 422 
\cite{wetterich}. 
The {\bf 210} representation of $SO(10)$ is contained in 
${\bf 16\times \overline{16}}={\bf 1}+{\bf 45}+{\bf 210}$, and 
so the 422 singlet in {\bf 210} comes from ${\bf (\bar{4},1,2)}
\times$ its conjugate and ${\bf (4,2,1)}\times$ its conjugate. 
One combination of these singlets gives the $SO(10)$ singlet, 
and the other the 422 singlet in {\bf 210}. The latter 
is the antisymmetric combination of these singlets and thus 
breaks the discrete $C$-parity which interchanges $SU(2)_L$ 
and $SU(2)_R$ and conjugates $SU(4)_c$ ($C$-parity, first 
found in Ref.~\cite{kibble}, was later called 
$D$-parity in Ref.~\cite{mohapatra}). This is clear since 
the $SO(10)$ singlet cannot break $C$, which belongs to 
$SO(10)$, and thus it is bound to be the symmetric 
combination. On the other hand, the 422 singlet in the 
{\bf 54}-plet of $SO(10)$ comes from the product 
${\bf 10\times 10}={\bf 1_{\rm s}}+
{\bf 45_{\rm a}}+{\bf 54_{\rm s}}$. Thus it originates 
from ${\bf (1,2,2)\times (1,2,2)}$ or ${\bf (6,1,1)\times 
(6,1,1)}$, which are both symmetric under $C$. One 
combination of them is the SO(10) singlet and the orthogonal 
combination is contained in {\bf 54}, and so the {\bf 54}-plet 
does not break the discrete symmetry $C$ \cite{kibble}. 

We choose here to employ a Higgs {\bf 210}-plet for the 
$SO(10)$ breaking 
to 422 so that no strings or subsequent walls bounded by 
strings are generated as in Ref.~\cite{kibble}. It is known 
\cite{magg} that the (-1,-1,-1) element of 422 coincides with 
the identity, and therefore three lines, one in each of the 
three groups 
between 1 and -1 constitute a closed loop, which corresponds 
to a magnetic monopole. We will now show that this monopole 
evolves to the Dirac monopole after the electroweak symmetry 
breaking. It carries one unit of magnetic charge as well as 
some color magnetic charge. (This conclusion appears to be in 
disagreement with Table III in Ref.~\cite{king2} where it is 
stated that the monopole is unstable.)

To make the analysis more transparent, we take the curve 
in $SU(4)_c$ along its $X\equiv(B-L)+2T^8_c/3$ generator, 
where $T^8_c={\rm diag}(1,1,-2)$ in $SU(3)_c$ and $B$ and 
$L$ are the baryon and lepton number operators respectively. 
This choice is certainly equivalent to taking the curve 
along the generator $B-L$ since color is unbroken. It is 
easy to see that $X={\rm diag}(1,1,-1,-1)$ in $SU(4)_c$ and 
the curve between 1 and -1 corresponds to a rotation by $\pi$ 
along this generator. In $SU(2)_L$ and $SU(2)_R$, we take 
rotations by $\pi$ along $T^3_L={\rm diag}(1,-1)$ and 
$T^3_R={\rm diag}(1,-1)$ respectively, and the overall loop
therefore  
corresponds to a rotation by $2\pi$ along $(B-L)/2+T^8_c/3+
T^3_L/2+T^3_R/2=Q+T^8_c/3$ ($Q$ is the electric charge 
operator). It is clear that this rotation brings us back 
to the identity element. Indeed, the group element 
$\exp(i2\pi T^8_c/3)=\exp(2i\pi/3)$ lies in the center of 
$SU(3)_c$ and $\exp(2i\pi Q)=\exp(4i\pi/3)$ acting on 
up-type quarks or 
$\exp(-2i\pi/3)$ acting on down-type quarks and so the 
combined rotation leads to the identity element. The 
magnetic monopole corresponding to a rotation by $2\pi$ 
along the generator $Q+T^8_c/3$ is exactly the 
Dirac magnetic monopole as previously shown in 
Ref.~\cite{daniel}. Note that, in this paper, the 
SM was embedded in $SU(5)$, but the 
argument holds for any compact group containing $SU(5)$. 
The Dirac magnetic monopole, along with the ordinary 
magnetic field, also carries color magnetic field. 

The next breaking of 422 to $SU(3)_c\times SU(2)_L
\times U(1)_{B-L}\times U(1)_R$ will generate monopoles 
corresponding to rotations by $2\pi$ in $SU(4)_c$ and 
$SU(2)_R$ along $X$ and $T^3_R$ respectively. This 
breaking can be achieved by the vacuum expectation value 
(VEV) of a Higgs {\bf 45}-plet of $SO(10)$ along its 
${\bf (1,1,3)}$ and ${\bf (15,1,1)}$ components. The 
${\bf (15,1,1)}$ VEV could also be taken from a Higgs
{\bf 210}-plet. 

We can further break these two $U(1)$'s by the VEV of the 
$\nu^c$-type SM singlet component in a Higgs
{\bf 16}-plet, which leaves $X+T^3_R$ unbroken ($\nu^c$ 
represents right-handed neutrinos). To find the broken 
generator which is perpendicular to $X+T^3_R$, we must 
define the GUT normalized generators
\beq
Q_X=\frac{1}{2\sqrt{2}}X, \quad Q_R=\frac{1}{2}T^3_R.
\eeq
Then the normalized unbroken and broken generators 
$\mathcal{U}$ and $\mathcal{B}$ are, respectively, 
\bea
\mathcal{U}&=&\frac{1}{\sqrt{3}}(\sqrt{2}Q_X+Q_R), 
\nonumber\\
\mathcal{B}&=&\frac{1}{\sqrt{3}}(Q_X-\sqrt{2}Q_R)=
\frac{1}{2\sqrt{6}}(X-2T^3_R).
\eea

\begin{figure}[t]
\centerline{\epsfig{file=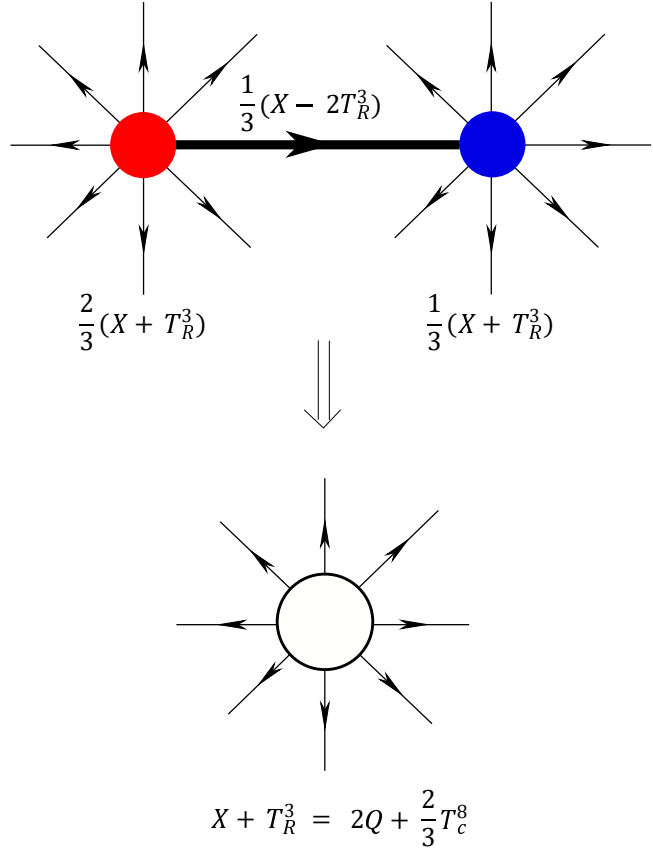,width=8.7cm}}
\caption{Emergence of (Schwinger) magnetic monopole with 
two units of Dirac charge from the symmetry breaking 
$SU(4)_c\times SU(2)_L\times SU(2)_R\to
SU(3)_c\times U(1)_{B-L}\times SU(2)_L\times U(1)_R\to
SU(3)_c\times SU(2)_L\times U(1)_Y$. This monopole also 
carries color magnetic charge. An $SU(4)_c$ (red) and an 
$SU(2)_R$ (blue) monopole are connected by a flux tube 
which pulls them together to form a Schwinger monopole.
The magnetic flux along the tube and the Coulomb magnetic 
fluxes of the monopoles are indicated. Intermediate mass 
monopoles such as this one may survive inflation.}
\label{fig:schwinger}
\end{figure}

The smallest broken generator with integral charges so 
that its periodicity is $2\pi$ is $X-2T^3_R$. A rotation 
along this generator by $2\pi/3$ is left unbroken by the 
VEV of the $\nu^c$-type Higgs. Therefore, the 
generated string contains magnetic flux corresponding to 
a rotation by $2\pi/3$ along $(X-2T^3_R)$.
The magnetic fluxes of an $SU(4)_c$ and an $SU(2)_R$
monopole have to be rearranged in tubes with flux 
$(X-2T^3_R)/3$ and Coulomb fluxes along the
unbroken generator $X+T^3_R$. To this end, 
an $SU(4)_c$ monopole, which carries a full flux 
along $X$, sends 1/3 of it to an $SU(2)_R$ monopole 
which carries a full $T^3_R$ flux. This latter monopole, 
in turn, sends 2/3 of its flux to the other one and thus a 
tube is generated between them which pulls them together. 
The rest of the fluxes are added together to give the 
Coulomb flux of a doubly charged (Schwinger) monopole -- 
see Fig.~\ref{fig:schwinger}. Note that the 1/3 of the $X$ 
flux sent from the $SU(4)_c$ monopole towards the $SU(2)_R$ 
monopole to contribute to the tube in between cannot 
terminate on it but emerges as Coulomb flux from it. The 
same is true for the 2/3 of $T^3_R$ flux sent from the 
$SU(2)_R$ monopole to the $SU(4)_c$ monopole. Finally, 
we have four fluxes (two corresponding to rotations by 
$4\pi/3$ and $2\pi/3$ along $X$, and two corresponding to  
rotations by $2\pi/3$ and $4\pi/3$ along $T^3_R$), combined 
together to emerge as Coulomb flux from the combined 
monopole. This monopole corresponds to a full ($2\pi$) 
rotation along $X+T^3_R$ and becomes a Schwinger monopole 
after the electroweak breaking. Needless to say the 
$SU(4)_c$ or $SU(2)_R$ monopoles can be connected by a 
string to their respective antimonopoles and annihilate.

Note that
\bea
& &\exp\left\{i\frac{2\pi}{3}(X-2T^3_R)\right\}
=\exp\left\{i\frac{2\pi}{3}(X+T^3_R)\right\}\times
\nonumber\\
& &\exp\left(-i2\pi T^3_R\right)=\exp\left\{i\frac{2\pi}
{3}(X+T^3_R)\right\},
\eea
and thus this unbroken element belongs to the unbroken
continuous subgroup, i.e. the SM group. 
Consequently, no unbroken discrete symmetry is left, 
which means that no topologically stable strings are 
produced since the first homotopy (fundamental) group
of the vacuum manifold
\bea
& &\pi_1\left(\frac{SO(10)}{SU(3)_c\times SU(2)_L\times 
U(1)_Y}\right)
\nonumber\\
& &=\pi_0\left(SU(3)_c\times SU(2)_L\times U(1)_Y\right)
=\{1\}.
\eea
We only have dumbbells \cite{dumbbells} which can 
transform into Schwinger monopoles. 

If we inflate away the 
$SU(4)_c$ and $SU(2)_R$ monopoles, we can have a 
network of topologically non-stable strings. After the 
electroweak breaking, the Higgs doublets $h_u$, $h_d$ 
($h_u$ couples to the up-type quarks and $h_d$ to the 
down-type ones) with $X=0$ and $T^3_R=1,-1$, $T^3_L=-1,1$ 
respectively develop VEVs. As we circle a string they get 
a phase $-4\pi/3$, $4\pi/3$ respectively. If we 
add to the string 1/3 of flux along $T^3_L$ so that the 
string corresponds to a rotation by $2\pi/3$ along 
$X-2T^3_R+T^3_L$, 
the phases of $h_u$ and $h_d$ change by $-2\pi$ and $+2\pi$ 
respectively around the string. Of course, this addition 
does not affect the $\nu^c$-type VEV of the Higgs 
{\bf 16}-plet and also adds the minimal necessary magnetic 
energy on the string. For definiteness, we will assume 
throughout that the magnetic energy dominates over the 
Higgs contribution to the string energy and so these strings 
are superconducting \cite{witten}. We obtain left-moving 
and right-moving fermionic zero modes along the string 
via $h_u$, $h_d$ which are the only Higgs fields 
coupling to quarks and charged leptons. Note that the 
$\nu^c$-type Higgs field couples only to right-handed 
neutrinos and thus does not contribute to superconductivity.

\begin{figure}[t]
\centerline{\epsfig{file=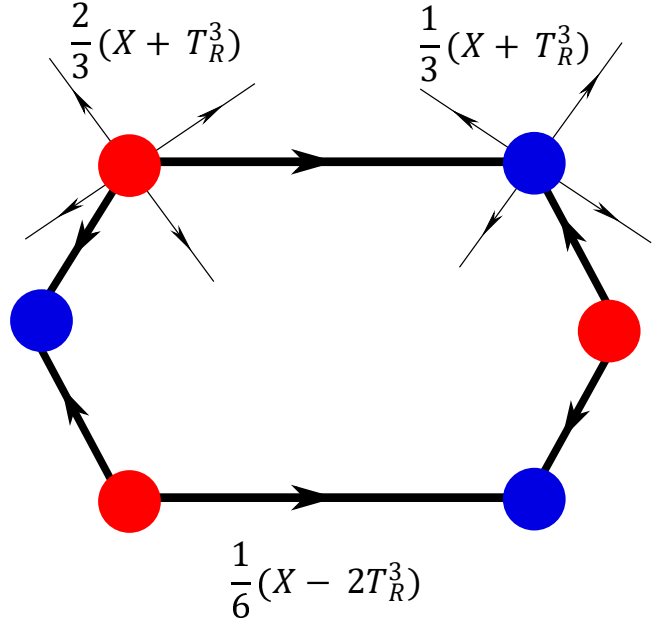,width=8.7cm}}
\caption{Necklace with $SU(4)_c$ and $SU(2)_R$ monopoles 
from the symmetry breaking $SU(4)_c\times SU(2)_L\times 
SU(2)_R\to SU(3)_c\times U(1)_{B-L}\times SU(2)_L\times 
U(1)_R\to SU(3)_c\times SU(2)_L\times U(1)_Y\times Z_2$, 
where the last step is achieved by a {\bf 126}-plet
of $SO(10)$. Notation as in Fig.~\ref{fig:schwinger}. 
We display explicitly only the Coulomb magnetic flux of 
two of the monopoles and the magnetic flux along one of 
the tubes. This necklace may survive inflation.}
\label{fig:necklace}
\end{figure}

Now suppose that we use the $\nu^c\nu^c$-type component of 
{\bf 126} to do the breaking of $X-2T^3_R$. In this case, 
a rotation by $2\pi/6$ along $X-2T^3_R$ leads to an unbroken 
element. This yields a string which contains magnetic flux 
corresponding to a rotation by $2\pi/6$ along $X$ minus flux 
corresponding to a rotation by $2\pi/3$ along $T^3_R$. An 
$SU(4)_c$ and an $SU(2)_R$ monopole are then connected by two
such strings with the remaining Coulomb flux in them being 
$(X+T^3_R)/3$ and $2(X+T^3_R)/3$. Now if one imagines opening 
up one of the two strings, one finds the two monopoles 
connected by one string and two ``loose'' strings emerging 
from the two monopole. One can then connect these latter 
strings to other similar monopole-string structures in series 
and form necklaces \cite{vachaspati} -- see 
Fig.~\ref{fig:necklace}. Note that pairs of $SU(4)_c$ and 
$SU(2)_R$ antimonopoles connected by a string can also 
participate in the necklace with the $SU(4)_c$ antimonopole 
connected either to an $SU(4)_c$ monopole or an $SU(2)_R$ 
antimonopole, and the $SU(2)_R$ antimonopole connected either 
to an $SU(2)_R$ monopole or $SU(4)_c$ antimonopole. Also both
tubes emerging from an $SU(4)_c$ monopole ($SU(2)_R$ 
antimonopole) can terminate on $SU(4)_c$ antimonopoles 
($SU(2)_R$ monopoles). We thus see that a variety of necklaces 
can appear with different arrangements of $SU(4)_c$ and 
$SU(2)_R$ monopoles and antimonopoles. 

The group element 
\bea
& &\exp\left\{i\frac{2\pi}{6}(X-2T^3_R)\right\}=
\exp\left\{i\frac{2\pi}{6}(X+T^3_R)\right\}\times 
\nonumber\\
& &\exp\left(-i\frac{2\pi}{2} T^3_R\right)=
\exp\left\{i\frac{2\pi}{6}(X+T^3_R)\right\}(1,1,-1),
\nonumber\\
\eea
which we obtain by circling one of these strings does not 
belong to the SM group since its action on the 
SM singlet $\nu^c$ yields $\exp(i\pi)=-1$. Moreover, since 
$(1,1,-1)=(-1,-1,1)$ and $SU(2)_L$ is unbroken at this 
stage, this element is equivalent to the generator of 
the $Z_2$ subgroup of $U(1)_{B-L}$ \cite{z2string}. Its 
square is then obviously equivalent to the identity, and
an extra $Z_2$ symmetry remains unbroken. Stable $Z_2$ 
strings without monopoles on them are also present. 
These strings, exactly like the ones in the necklaces above, 
correspond to a rotation by $2\pi/6$ along $X-2T^3_R$ and 
are not oriented. The necklaces are themselves $Z_2$ 
strings too. 

Next let us see what happens after the electroweak symmetry 
breaking. Recall that the Higgs doublets $h_u$, $h_d$ have 
$X=0$ and $T^3_R=1,-1$, $T^3_L=-1,1$ respectively. As we 
go around the string they acquire a phase $-2\pi/3$, $+2\pi/3$ 
respectively. If we add to the string -1/3 of flux along 
$T^3_L$ such that the string corresponds to a rotation by 
$2\pi/6$ 
along $X-2T^3_R-2T^3_L$, $h_u$, $h_d$ remain constant around 
the string. Of course, this addition does not affect the 
$\nu^c\nu^c$-type VEV of {\bf 126} and also adds the minimal 
necessary magnetic energy along the string. Thus, only the 
$\nu^c\nu^c$-type component of {\bf 126} changes phase around 
the string. But this couples only to right-handed neutrinos 
and so these strings are not superconducting.

 \begin{figure}[t]
\centerline{\epsfig{file=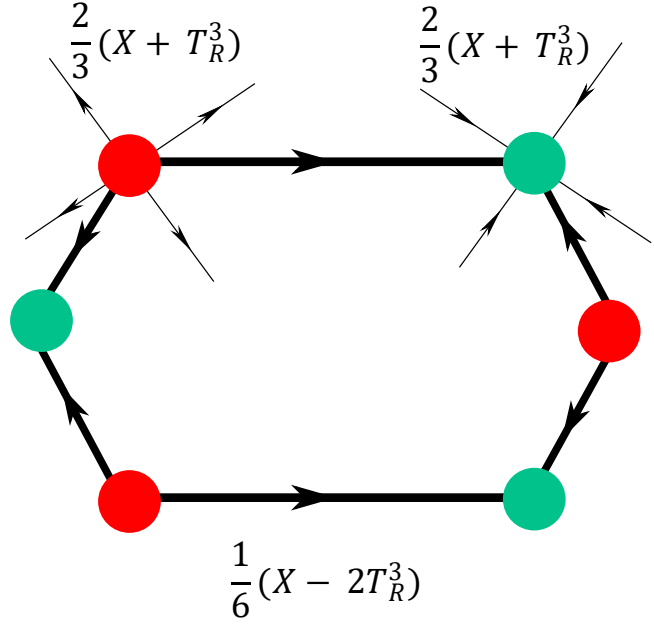,width=8.7cm}}
\caption{Necklace with $SU(4)_c$ monopoles (red) and 
antimonopoles (green) from the symmetry breaking $SO(10)
\to SU(4)_c\times SU(2)_L\times U(1)_R\to SU(3)_c\times 
U(1)_{B-L}\times SU(2)_L\times U(1)_R\to SU(3)_c\times 
SU(2)_L\times U(1)_Y\times Z_2$, where the last step is 
achieved by a {\bf 126}-plet of $SO(10)$. We assume that 
the monopoles from the first step of symmetry breaking are 
inflated away. We display explicitly only the Coulomb 
magnetic flux of one monopole and one antimonopole and the 
magnetic flux along one of the tubes. This necklace may 
survive inflation.}
\label{fig:simplenecklace}
\end{figure}

For an example of a monopole-antimonopole necklace formed 
with a $Z_2$ string, consider the following $SO(10)$ breaking 
pattern: $SO(10)\to SU(4)_c\times SU(2)_L\times U(1)_R\to 
SU(3)_c\times U(1)_{B-L}\times SU(2)_L\times U(1)_R\to SU(3)_c
\times SU(2)_L\times U(1)_Y\times Z_2$. The first 
breaking, achieved by the VEVs of a {\bf 210}-plet and a 
{\bf 45}-plet along their {\bf (1,1,1)} and {\bf (1,1,3)} 
components respectively, produces a GUT monopole with one unit 
of Dirac magnetic charge, which presumably is inflated away.
Of course, multiply charged monopoles may also be produced and 
inflated away. In particular, the doubly charged monopole 
coincides with the $SU(2)_R$ monopole we mentioned above since
the corresponding loops in $SU(4)_c$ and $SU(2)_L$ are 
homotopically trivial. The second breaking, achieved by the 
VEV of the {\bf (15,1,1)} component of a Higgs {\bf 45}-plet, 
yields an intermediate scale $SU(4)_c$ monopole which carries 
both $SU(3)_c$ and $U(1)_{B-L}$ magnetic fluxes. The last 
breaking is done by the $\nu^c\nu^c$-type component of a Higgs
{\bf 126}-plet and the $SU(4)_c$ monopoles can form, together 
with the antimonopoles, a necklace tied together by a $Z_2$ 
string. Namely, an $SU(4)_c$ monopole, which carries a full
magnetic flux along $X$, rearranges its magnetic field to form
two tubes with flux $(X-2T^3_R)/6$ and a Coulomb field around 
it with flux $2(X+T^3_R)/3$. Since the $SU(2)_R$ monopoles are
inflated away in this case, these tubes can only terminate on
$SU(4)_c$ antimonopoles -- see Fig.~\ref{fig:simplenecklace}. 
   
\section{$E_6$ breaking via $SU(3)_c\times SU(3)_L\times 
SU(3)_R$}
\label{sec:333}      
 
$E_6$ can break to the trinification group $SU(3)_c\times 
SU(3)_L\times SU(3)_R$ (333, for short) by the VEV of a 
Higgs {\bf 650}-plet, 
which contains two 333 singlets. One of them breaks 
$C$-parity, but the other one does not. Note that the 
symmetry $C$, in this case, exchanges $SU(3)_L$ and 
$SU(3)_R$ and conjugates the representation, in which case
${\bf (1,\bar{3},3)}$ goes to itself, while {\bf (3,3,1)} 
and ${\bf (\bar{3},1,\bar{3})}$ are interchanged. There are
three 333 singlets in the product
\beq
{\bf 27}\times {\bf \overline{27}}={\bf 1}+{\bf 78}+{\bf 650}.
\eeq
They are the ${\bf (1,\bar{3},3)\times (1,3,\bar{3})}$, 
${\bf (3,3,1)\times (\bar{3},\bar{3},1)}$, 
and ${\bf (\bar{3},1,\bar{3})\times (3,1,3)}$. 
The sum of these three singlets gives the $E_6$ singlet. The 
other two orthogonal combinations are in {\bf 650} since
{\bf 78} has no 333 singlet. They could be
\bea
& &2{\bf (1,\bar{3},3)(1,3,\bar{3})}-
{\bf (3,3,1)(\bar{3},\bar{3},1)}
\nonumber\\
& &-{\bf (\bar{3},1,\bar{3})(3,1,3)},
\nonumber\\
& &{\bf (3,3,1)(\bar{3},\bar{3},1)}-
{\bf (\bar{3},1,\bar{3})(3,1,3)}.
\eea
The latter violates $C$. Both these singlets can acquire VEVs 
and thus $C$ will be broken, and we expect that no strings or 
walls bounded by strings \cite{kibble} associated with $C$ are 
generated. 

The fundamental representation of $E_6$ is 
\beq
{\bf 27}={\bf (1,\bar{3},3)}+{\bf (3,3,1)}+
{\bf (\bar{3},1,\bar{3})}\equiv\lambda+Q+Q^c, 
\eeq
where
\beq
\lambda= 
\begin{pmatrix}
h_u & e^c
\\
&
\\
h_d & \nu^c 
\\
&
\\  
l  & N
\end{pmatrix}        
\eeq         
with the rows being ${\bf \bar{3}}$'s of $SU(3)_L$ and the 
columns {\bf 3}'s of $SU(3)_R$, and
\beq
Q= 
\begin{pmatrix}
q
\\
&
\\
g
\end{pmatrix} \quad {\rm and}\quad
Q^c=
\begin{pmatrix}
u^c, &d^c, &g^c 
\end{pmatrix},   
\eeq
which are an $SU(3)_L$ triplet and an $SU(3)_R$ antitriplet 
respectively. 

One can very easily verify that the element $c=\left(\exp(i2\pi/3), 
\exp(-i2\pi/3), \exp(-i2\pi/3)\right)$ of the unbroken 
trinification subgroup $H$ coincides with the identity 
element as it acts like the identity on the {\bf 27}-plet 
and, consequently, on all the representations of $E_6$. The 
generator of the second homotopy group $\pi_2(E_6/H)=
\pi_1(H)=Z_3$ of the vacuum manifold $E_6/H$ is then 
a loop that connects (1,1,1) with $c$, i.e. three curves in 
the three $SU(3)$'s from 1 to $\exp(i2\pi/3)$, or 1 to 
$\exp(-i2\pi/3)$, or 1 to $\exp(-i2\pi/3)$ respectively. 
Obviously, the third power of this loop is homotopically 
trivial, and the breaking $E_6\to {\rm 333}$ therefore 
generates $Z_3$ magnetic monopoles.

In order to understand the structure of these $Z_3$ monopoles, 
we define the generators $T^8_L={\rm diag}(1,1,-2)$, 
$T^3_L={\rm diag}(1,-1,0)$ of $SU(3)_L$ and 
$T^8_R={\rm diag}(1,1,-2)$, $T^3_R={\rm diag}(1,-1,0)$ of 
$SU(3)_R$. Note that we use integer elements in these 
definitions so that a full rotation by $2\pi$ along these 
generators closes a circle. We see that 
$(1/6)T^8_L+(1/2)T^3_L={\rm diag}(2/3,-1/3,-1/3)$, and a 
rotation by $2\pi$ along this generator brings us from 1 to 
$\exp(-i2\pi/3)$ in $SU(3)_L$. Similarly, a rotation by $2\pi$ 
along the generator $(1/6)T^8_R+(1/2)T^3_R$ interpolates 
between 1 and $\exp(-i2\pi/3)$ in $SU(3)_R$. In $SU(3)_c$, we 
take a rotation by $2\pi/3$ along $T^8_c={\rm diag}(1,1,-2)$, 
which leads from 1 to the element $\exp(i2\pi/3)$. The 
generator of the first homotopy (fundamental) group 
$\pi_1(H)=Z_3$ of $H$ can be represented by a $2\pi$ rotation 
along the generator
\beq
\frac{1}{3}T^8_c+\frac{1}{6}T^8_L+\frac{1}{2}T^3_L+
\frac{1}{6}T^8_R+\frac{1}{2}T^3_R.
\label{generator}
\eeq
It is easy to check that 
\beq
\frac{1}{6}T^8_L+\frac{1}{2}T^3_L+\frac{1}{6}T^8_R+\frac{1}{2}T^3_R
=Y+\frac{1}{2}T^3_L=Q, 
\eeq
the electric charge operator, by applying it on the
various states in {\bf 27} ($Y$ is the weak hypercharge).
Finally, we see that the generator of $\pi_1(H)$ is a rotation by 
$2\pi$ along $T^8_c/3+Q$, exactly as in the $SO(10)$ case. As a 
consequence, the $Z_3$ monopole in $E_6$, similarly to the $Z_2$ 
monopole in $SO(10)$, carries one (Dirac) unit of ordinary magnetic 
flux or charge as well as color magnetic flux corresponding to the 
generator of the center of $SU(3)_c$. As shown in Ref.~\cite{daniel} 
this is the ordinary Dirac monopole also carrying color magnetic 
charge.

We can further break 333 to $SU(3)_c\times SU(2)_L\times  SU(2)_R
\times U(1)_{B-L}$ (${\rm 3221}_{B-L}$, for short) by giving a VEV to 
the $N$-type component of ${\bf (1,\bar{3},3)}$ in a Higgs 
{\bf 27}-plet. The generator in Eq.~(\ref{generator}) remains in 
the unbroken subalgebra since
\beq
\frac{1}{6}(T^8_L+T^8_R)=\frac{1}{2}(B-L).
\eeq
The orthogonal broken generator is 
\beq
T^8_L-T^8_R,
\eeq
but a rotation by $2\pi/4$ along this generator leaves $N$ 
invariant and thus remains unbroken. Adding to this a 
rotation by $2\pi/4$ along the unbroken generator $T^8_L+T^8_R$, 
we get an equivalent rotation by $2\pi/2$ along $T^8_L$. This 
rotation corresponds to the group element $\exp(i2\pi T^8_L/2)=
{\rm diag}(-1,-1,1)$ in $SU(3)_L$, which belongs to the 
continuous part of the unbroken subgroup ${\rm 3221}_{B-L}$. 
This means that no additional discrete symmetries are left 
unbroken. In other words, the unbroken subgroup is precisely
${\rm 3221}_{B-L}$. 

The second homotopy group of the vacuum manifold $\pi_2({\rm 333}/
{\rm 3221}_{B-L})=\pi_1({\rm 3221}_{B-L})_{\rm 333}$, which means 
that it consists of the 
homotopically non-trivial loops in ${\rm 3221}_{B-L}$ which are trivial 
in 333. The minimal loop is a $6\pi$ rotation along the 
generator $T^8_c/3+Q$, and so the loop in $SU(3)_c$ becomes 
homotopically trivial and can be removed. Only the rotation along 
$Q$ by $6\pi$ remains, which corresponds to a monopole with triple 
the ordinary magnetic charge and no color magnetic flux at all.

The subsequent breaking of $SU(2)_R\times U(1)_{B-L}$ to 
$U(1)_Y$ does not generate any new topological objects provided 
that it is done by an $SU(2)_R$ Higgs doublet, analogous to 
the electroweak breaking -- for a detailed explanation of this 
fact, see Ref.~\cite{trotta}. This breaking can be achieved by 
the VEV of a Higgs {\bf 27} along the $\nu^c$-type component of it. 
This belongs to an $SU(2)_R$ doublet with $B-L=1$ and generates no 
topological defects. 

We could alternatively use for the spontaneous breaking of 
$SU(2)_R\times U(1)_{B-L}$ to $U(1)_Y$ a Higgs 
$\overline{{\bf 351}'}$ 
(contained in ${\bf 27}\times {\bf 27})$ with a VEV along its 
${\bf (1,\bar{6},6)}$ component. In particular, we take the 
$SU(2)_R$ {\bf 3}-plet in the $SU(3)_R$ {\bf 6}-plet with 
$T^8_R=2$ and the $SU(2)_L$ singlet in the ${\bf \bar{6}}$ of 
$SU(3)_L$ with $T^8_L=4$. This is an $SU(2)_R$ triplet with 
$B-L=(T^8_R+T^8_L)/3=2$ and has the quantum numbers of 
$\nu^c\nu^c$. It thus leaves the $Z_2$ subgroup of $U(1)_{B-L}$ 
unbroken. So, in this case, in addition to the two types of 
monopoles, we have $Z_2$ strings as in Ref.~\cite{z2string}. 
However, 
there are no necklaces in this case. Note that the electroweak
Higgs doublets have zero $B-L$ and thus remain constant around the 
string. The string is not superconducting just as the 
string from the $Z_2$ subgroup of $U(1)_{B-L}$ in the previous 
section.

\section{$E_6$ breaking via $SO(10)\times U(1)_\psi$} 
\label{sec:psi}

Let us now turn to the case of $E_6\to SO(10)\times U(1)_\psi$. 
This can be achieved by a Higgs {\bf 78}-plet. We can further 
break $SO(10)\times U(1)_\psi \to SU(5)\times U(1)_{\psi'}$, 
where $\psi'=(\chi+5\psi)/4$, with $\chi$ corresponding to the 
$SU(5)\times U(1)_\chi$ subgroup of $SO(10)$. The $\psi$ charges 
for the $SO(10)$ components of the {\bf 27}-plet are given in 
parentheses

\beq
{\bf 27}={\bf 1}(4)+{\bf 10}(-2)+{\bf 16}(1),
\eeq
while the $(\chi,\psi)$ charges of its $SU(5)$ parts are
\bea
{\bf 27}&=&{\bf 1}(0,4)+{\bf 5}(2,-2)+{\bf \bar{5}}(-2,-2)
\nonumber\\
& &+{\bf 1}(-5,1)+{\bf \bar{5}}(3,1)+{\bf 10}(-1,1).
\eea
The breaking of $SO(10)\times U(1)_\psi\to SU(5)\times 
U(1)_{\psi'}$ is achieved by the VEV of {\bf 1}(-5,1), and 
thus the unbroken $U(1)$ corresponds to 
$\psi'=(\chi+5\psi)/4$ with the $\psi'$ charges given as
\beq
{\bf 27}={\bf 1}(5)+{\bf 5}(-2)+{\bf \bar{5}}(-3)+{\bf 1}(0)+
{\bf \bar{5}}(2)+{\bf 10}(1).
\eeq 
Note that in the definition of $\psi'$ we divided by 4 so 
that the $\psi'$ charges are the minimal integer ones (as 
the $\chi$ and $\psi$ charges), so that the periodicity of 
$U(1)_{\psi'}$ is $2\pi$.

The $U(1)_\psi$ intersects with $SO(10)$ in its $Z_4$ center. 
This center is generated by 
$-i\Gamma^{10}$, where 
\bea
\Gamma^{10}&=&i^5\Gamma^0\Gamma^3\Gamma^1\Gamma^2\Gamma^4
\Gamma^5\Gamma^7\Gamma^8\Gamma^6\Gamma^9
\nonumber\\
&=&\sigma^{03}\sigma^{12}\sigma^{45}\sigma^{78}\sigma^{69}
\eea
is the chirality operator in ten Euclidean dimensions. Here 
we use the notation of Ref.~\cite{yasue}, which follows the 
notation of Ref.~\cite{georgi}. The $SO(10)$ ${\bf 16}$-plet
(${\bf 1}
+{\bf \bar{5}}+{\bf 10}$) is of negative chirality and so the 
$SU(5)$ singlet {\bf 1} corresponds to all $\sigma$'s 
being -1, the ${\bf \bar{5}}$ to only one of them being -1 and all 
others +1, and the {\bf 10} to three of them being -1 and the rest +1. 
So under $-i\Gamma^{10}$, ${\bf 16} \to i {\bf 16}$ 
and, consequently, ${\bf 10} \to -{\bf 10}$ and ${\bf 1} \to {\bf 1}$. 
Now 
\bea
i\Gamma^{10}&=&i\sigma^{03}i\sigma^{12}i\sigma^{45}i\sigma^{78}
i\sigma^{69}
\nonumber\\
&=&\exp\left\{\frac{i\pi}{2}(\sigma^{03}+\sigma^{12}+\sigma^{45}+
\sigma^{78}+\sigma^{69})\right\}.
\nonumber\\
\eea  
It is easy to see that the sum of $\sigma$'s coincides with the
$\chi$ charge since it gives -5 for the $SU(5)$ singlet 
{\bf 1}, -1 for the {\bf 10}, and 3 for ${\bf \bar{5}}$. 
So the generator $-i\Gamma^{10}$ of the center of $SO(10)$ 
lies in $U(1)_\chi$ and corresponds to a rotation by 
$-2\pi/4$ along it.

Also, a rotation by $2\pi/4$ along $\psi$ acts on the $SO(10)$ 
representations as follows: ${\bf 16}\to i{\bf 16}$, 
${\bf 10}\to -{\bf 10}$, ${\bf 1}\to {\bf 1}$ and thus 
coincides with $-i\Gamma^{10}$. A rotation by $2\pi/4$ 
along $\psi$ together with a rotation by $2\pi/4$ along $\chi$ 
is a closed loop in $SO(10)\times U(1)_\psi$. This corresponds 
to the smallest charge magnetic monopole generated by the breaking 
$E_6\to SO(10)\times U(1)_\psi$. It has 1/4 of magnetic flux 
along $\psi$ and also an $SO(10)$ flux corresponding to the inverse 
generator of its center $i\Gamma^{10}$. A fourfold monopole, 
i.e. a monopole with magnetic flux equal to four times the flux 
of the minimal charged monopole, corresponds to a full ($2\pi$) 
rotation along $\psi$, since a full rotation along $\chi$ is 
homotopically trivial in $SO(10)$. 

Instead of using rotations along $\psi$ and $\chi$, it is 
more transparent to use rotation along $\psi$ and $\psi'$. 
Note that $\psi'=(\chi+\psi)/4+\psi$. A rotation by $2\pi$ 
along $(\chi+\psi)/4$ corresponds to the identity as we 
have just seen, and a rotation by $2\pi$ along $\psi$ again 
is the identity as one can see from the $\psi$ charges. The 
$\psi$ direction has no common elements with the center of 
$SU(5)$ since the $\psi$ charges of the $SU(5)$ singlets are 
4 and 1. However, the $\psi'$ direction has elements which 
coincide with the center of $SU(5)$. Namely, $\exp(i2\pi/5)$ 
in $U(1)_{\psi'}$ coincides with the element 
$\exp(-i2\pi\bar{Y}/5)$ of the center of $SU(5)$ with 
$\bar{Y}={\rm diag}(2,2,2,-3-3)$ in $SU(5)$. It is known 
\cite{langacker} that $\chi$ and $\psi$ correspond to the 
following GUT normalized generators 
\beq
Q_\chi=\frac{\chi}{2\sqrt{10}},\quad Q_\psi=
\frac{\psi}{2\sqrt{6}}.
\eeq
Then the normalized generator for $\psi'$ is
\beq
Q_{\psi'}=\frac{1}{4}\left(Q_\chi+\sqrt{15}Q_\psi\right)=
\frac{\psi'}{2\sqrt{10}},
\eeq
and the orthogonal generator is
\beq
Q_{\chi'}= \frac{1}{4}\left(
\sqrt{15}Q_\chi-Q_\psi\right)=\frac{\chi'}{2\sqrt{6}},
\eeq
with $\chi'=(3\chi-\psi)/4$. The $\chi'$ charges are 
\beq
{\bf 27}={\bf 1}(-1)+{\bf 5}(2)+{\bf \bar{5}}(-1)+
{\bf 1}(-4)+{\bf \bar{5}}(2)+{\bf 10}(-1),
\eeq
such that the full
rotation along $\chi'$ is a $2\pi$ rotation. 

Note that the $\chi'$ direction has no common elements 
with $SU(5)$ since the charges of the $SU(5)$ singlets in 
{\bf 27} are -1 and -4. However, the $Z_4$ subgroups from 
$\psi'$ and $\chi'$ coincide. Namely, a rotation 
by $2\pi/4$ along $\psi'$ together with a rotation by 
$2\pi/4$ along $\chi'$ lead to the identity element as one 
can see from the $\psi'$, $\chi'$ charges. It is, as we 
will see, more convenient to use the orthogonal generators 
$\psi'$, $\chi'$ rather than $\psi$, $\chi$.

What happens in the next breaking to $SU(5)\times 
U(1)_{\psi'}$ by the $\nu^c$-type component of a Higgs 
{\bf 27}-plet, i.e. the singlet in its $SO(10)$ 
{\bf 16}-plet? The $U(1)_{\psi'}$ symmetry remains, of 
course, unbroken, 
but the orthogonal $U(1)_{\chi'}$ breaks to its $Z_4$ 
subgroup since the $\chi'$ charge of the $\nu^c$-type 
component is 
-4. However, this $Z_4$ belongs to $U(1)_{\psi'}$ and thus 
the unbroken subgroup is just $SU(5)\times U(1)_{\psi'}$, 
which is connected, i.e. its zeroth homotopy group 
$\pi_0(SU(5)\times U(1)_{\psi'})=\{1\}$. Therefore, the 
first homotopy (fundamental) group of the vacuum manifold
\beq
\pi_1\left(\frac{E_6}{SU(5)\times U(1)_{\psi'}}\right)
=\pi_0(SU(5)\times U(1)_{\psi'})=\{1\},
\eeq
and no stable strings will appear at this stage. 

How about the previous monopole with magnetic flux 
$(\chi+\psi)/4=(\psi'+\chi')/4$? As $U(1)_{\chi'}$ breaks to 
its $Z_4$ subgroup (which belongs to $U(1)_{\psi'}$ too), the 
$\chi'/4$ flux of the monopole is confined to a tube and 
connects it to an antimonopole. We therefore obtain unstable 
dumbbells 
\cite{dumbbells} which disappear. The Coulomb $\psi'/4$ and 
$-\psi'/4$ fluxes of the monopole and antimonopole, of course, 
cancel each other. However, new monopoles appear which carry 
$U(1)_{\psi'}$ flux. Indeed, since the $Z_5$ subgroup of 
$U(1)_{\psi'}$ belongs to $SU(5)$, as we showed above, these 
monopoles correspond to a rotation by $2\pi/5$ along $\psi'$ 
and also carry  $SU(5)$ flux corresponding to the element 
$\exp(i2\pi\bar{Y}/5)$ of the center of $SU(5)$. We are 
left at this stage only with these $\psi'$ monopoles.
Next we can break $U(1)_{\psi'}$ by the $SO(10)$ singlet in 
a Higgs {\bf 27}-plet which, of course, leaves unbroken its 
$Z_5$ subgroup contained in the center of $SU(5)$. Then 
strings are formed with flux corresponding to a rotation by 
$2\pi/5$ along $\psi'$ which connect the $\psi'$ monopoles to 
antimonopoles leading them to annihilate. Thus, no topological 
defects survive at the end. From this point on, the story 
proceeds as usual with the breaking of $SU(5)$. 

It is interesting to note that we could inflate away the 
monopoles with magnetic flux $(\psi'+\chi')/4$ and obtain a 
network of cosmic strings with magnetic flux $\chi'/4$, which 
are not topologically stable. At the breaking of $U(1)_{\psi'}$ 
by the $N$-type component of the Higgs {\bf 27}, a $\psi'/20$ 
magnetic flux is added along these strings in order for the 
phase of 
$N$ to remain constant around them. This addition certainly 
corresponds to the minimal necessary increase of the magnetic 
energy of the string. At the electroweak breaking, the phases 
of the Higgs fields $h_u$ and $h_d$, which have $\chi'=2,-1$ 
and $\psi'=-2,-3$ respectively, change around the string by 
$4\pi/5$ and $-4\pi/5$ respectively. If we minimally 
add to the string -2/5 of flux along $2Y$, the VEVs of 
$h_u$, $h_d$ remain constant around it. Only the VEV of the 
$\nu^c$-type Higgs field changes its phase by $-2\pi$ around 
the string.  However, this string is not superconducting. 
Indeed, the up-type quark masses originate from the VEV of 
$h_u$ which remains constant around the string, and thus no 
transverse zero modes are generated along 
the string. The down-type quarks and charged lepton, although 
$h_d$ also remains constant, could generate zero modes since 
the $\nu^c$-type VEV contributes to their masses. 

Recall that $d^c$-type quarks ($e$-type charged leptons) exist 
not only in the ${\bf \bar{5}}$ in the $SO(10)$ {\bf 16}-plet, 
but also 
in the ${\bf \bar{5}}$ in the $SO(10)$ {\bf 10}-plet, which we 
call $D^c$ ($E$). Also, $d$-type quarks ($e^c$-type charged 
leptons) exist not only in the {\bf 10} in the $SO(10)$ 
{\bf 16}-plet but also in the {\bf 5} in the $SO(10)$ 
{\bf 10}-plet, which we call $D$ ($E^c$). Then the masses of 
the down-type quarks can be schematically written as
\beq
\mathcal{M}_d=
\begin{pmatrix}
D^c, & d^c
\end{pmatrix}
\begin{pmatrix} 
N, & \alpha_{ij} h_d\\
& \\
\nu^c, & h_d
\end{pmatrix}
\begin{pmatrix}
D \\
&\\
d
\end{pmatrix},
\label{matrix}
\eeq      
where the mass matrix is given in terms of four $3\times 3$
blocks. Three of them are of the order of the VEVs of $N$, 
$\nu^c$, and $h_d$ as indicated with constant unsuppressed 
coefficients. The fourth is proportional to the VEV of $h_d$ 
but multiplied by coefficients $\alpha_{ij}$ ($i,j=1,2,3$) 
which are suppressed by powers of $m_{\rm P}$, the Planck 
mass. This is due to fact that a direct trilinear Yukawa 
coupling is forbidden, in this case, by $U(1)_{\chi'}$ and 
$U(1)_{\psi'}$. The coefficients $\alpha_{ij}$ must then 
necessarily contain $U(1)_{\chi'}$ and $U(1)_{\psi'}$ 
violating SM singlet VEVs, i.e. $\vev{N}$, $\vev{\nu^c}$. 

We can now apply a 
theorem given in Ref.~\cite{ganoulis} which says that, if 
a particular mass matrix element remains constant around the 
string, we can remove from the mass matrix the row and the 
column that contain it when calculating the number of 
transverse zero modes. In our case $N$ and $h_d$ remain 
unaltered around the string, so all rows and columns can be 
removed and no zero modes appear. We see that the fact that 
$\nu^c$ changes phase around the string does not generate 
zero modes in this case. A very similar analysis can be done 
for the charged leptons by replacing $D^c$, $d^c$, $D$, $d$ 
in Eq.~(\ref{matrix}) by $E$, $e$, $E^c$, $e^c$ respectively. 
We conclude that these strings are not superconducting. 

We could also inflate away the monopoles with $\psi'/5$ 
and $SU(5)$ flux to get a network of strings with magnetic 
flux $\psi'/5$. Recall that the phase of $N$ changes by 
$2\pi$ around such a string, while $\nu^c$ remains 
constant. The phases of the electroweak doublets $h_u$, 
$h_d$ change by $(-2/5)2\pi$, $(-3/5)2\pi$ respectively. 
Adding minimally on the string 2/5 of flux along $2Y$, we 
then see that $h_u$ remains constant around the string, 
while the phase of $h_d$ changes by $-2\pi$. Again, 
we have no zero modes from the up-quark sector. For the 
down-quark and charged lepton sectors, we can write 
mass matrices similar to the one in Eq.~(\ref{matrix}).
Then, we can remove the rows and columns which contain 
elements proportional to $\nu^c$ which leaves the 
$3\times 3$ matrix which is proportional to $\alpha_{ij} 
h_d$. This matrix also does not change phase around the 
string as one can see from the various charges of the 
product $D^cd$. Thus, no transverse zero modes exist 
and these strings are also not superconducting.


Now if the breaking to $SU(5)\times U(1)_{\psi'}$ is 
achieved by the $\nu^c\nu^c$-type component of 
$\overline{{\bf 351}'}$, the $Z_8$ subgroup of $U(1)_{\chi'}$ 
remains unbroken. But the $Z_4$ subgroup of it is in 
$U(1)_{\psi'}$, so actually the unbroken subgroup is $SU(5)
\times U(1)_{\psi'}\times Z_2$. As a consequence, $Z_2$ 
strings are formed with flux $\chi'/8$. Note that these are 
$Z_2$ strings, i.e. the string and antistring coincide 
(they are not oriented). In this case, the $\chi'/4$ flux 
of the monopole with total flux $(\chi'+\psi')/4$ splits 
into two tubes, each with flux $\chi'/8$. The monopoles can 
then be connected to form necklaces which are $Z_2$ strings 
themselves. We can also have simple $Z_2$ strings with flux 
$\chi'/8$ without monopoles on them since
\bea
& &\pi_1\left(\frac{E_6}{SU(5)\times U(1)_{\psi'}\times Z_2}
\right)
\nonumber\\
& &=\pi_0(SU(5)\times U(1)_{\psi'}\times Z_2)=Z_2.
\label{z2}
\eea
Needless to say the $\psi'$ monopoles will also appear at 
this stage as before. 


We can further use the $SO(10)$ singlet in a Higgs 
{\bf 27}-plet 
to break $U(1)_{\psi'}$. Again, we obtain flux tubes 
carrying $\psi'/5$ magnetic flux which connect the $\psi'$ 
monopoles and antimonopoles and lead them to annihilation. 
If we inflate these 
$\psi'$ monopoles we obtain a network of non-superconducting 
strings with flux $\psi'/5$ as we have seen above. The VEV 
of the $N$-type component does not break the extra $Z_2$ in 
Eq.~(\ref{z2}), but merely rotates it. Actually, if we add 
1/40 of the flux along $\psi'$ on the string with flux 
$\chi'/8$, the phase of the $N$-type component remains 
unchanged around it. The electroweak doublets $h_u$, $h_d$ 
have $\chi'=2,-1$ and $\psi'=-2,-3$ respectively and thus, 
around the string, their phases change by $-2\pi/5$, 
$+2\pi/5$ respectively. If we minimally add to the 
string 1/5 of flux along $2Y$, the VEVs of $h_u$, $h_d$ 
remain constant around it. In summary, these strings and 
necklaces survive even after the electroweak breaking but 
they are not superconducting. 


\section{Primordial Monopoles, Strings, and Gravity Waves}
\label{sec:inf} 

As previously mentioned primordial monopoles and strings 
can survive inflation in realistic models. Consider, for 
instance, the breaking of $SO(10)$ to the SM 
via the 422 subgroup, such that the $Z_2$ subgroup of the 
center of $SO(10)$ remains unbroken. Assume that inflation 
is driven by an $SO(10)$ singlet scalar field with a Higgs 
or Coleman-Weinberg potential and with minimal coupling 
to gravity \cite{extended,Shafi:1983bd}. This model 
predicts that the tensor-to-scalar ratio $r\gtrsim 0.02$ 
\cite{Shafi:2006cs}. In other words, the Hubble parameter 
$H$ during observable inflation is estimated to 
be of order $10^{13}-10^{14}~{\rm GeV}$, which has important 
implications for primordial monopoles and strings. The GUT 
monopoles produced during the breaking of $SO(10)$ to 422 
are inflated away, but the intermediate mass monopoles from 
422 breaking at $M_{\rm I}$ may survive inflation if 
$M_{\rm I}\sim H$. In practice, one needs about 
$23-25$ $e$-foldings for adequate suppression and  
still leave an observable number density of these intermediate 
mass monopoles \cite{extended,senoguz} -- see below. By the 
same token the intermediate scale $Z_2$ cosmic strings which 
are produced during the breaking of 422 to the SM can also 
survive the inflationary epoch. In the case the $Z_2$ center
of $SO(10)$ is broken, we do not have topologically stable
strings. However, the monopoles produced during the breaking 
of 422 are connected, in the next stage of symmetry breaking,
by topologically non-stable strings. The monopole-string 
system eventually decays by emitting gravity waves which may 
be detectable by future experiments (see below). 
   
Regarding $E_6$, as we have shown, the symmetry breaking $E_6 
\to {\rm 333}$ yields a superheavy GUT monopole which is inflated 
away, at least in the inflationary scenarios we have mentioned 
here. However, analogous to the $SO(10)$ case, the triply 
charged intermediate mass ($\sim 10^{14}~{\rm GeV}$) monopoles 
from the breaking of 333 to the SM may be present 
at an observable level in our galaxy. Other realistic examples 
of intermediate scale monopoles, strings and composite objects 
that survive an inflationary scenario can be readily constructed.

We will now give some details concerning the production and 
evolution of monopoles and cosmic strings as well as 
the gravity waves generated by topologically stable
or unstable strings. The mean distance between topological 
defects (monopoles or strings)
at production is estimated to be $\sim H^{-1}$. During inflation it 
acquires an extra factor $e^\eta$ with $\eta$ being the number of 
$e$-foldings following the generation of the defects. During the 
subsequent inflaton oscillations this distance is multiplied by a 
factor $(t_{\rm r}/\tau)^{2/3}$ with $t_{\rm r}$ being the reheat 
time and $\tau$ the rollover time, and from reheating until the 
present time by another factor $T_{\rm r}/T_0$ ($T_0$
is the present temperature). So all together the mean distance 
between defects becomes
\begin{equation}
H^{-1} e^\eta \left(\frac{t_{\rm r}}{\tau}\right)^{2/3}\frac{T_r}
{T_0}\,\cdot
\label{distance}
\end{equation}
For $T_{\rm r}\simeq 10^9~{\rm GeV}$ and for the SM 
spectrum, 
$t_{\rm r}
\simeq 1~{\rm GeV}^{-1}$. For the defects to enter into the horizon 
until today, their present mean distance in Eq.~(\ref{distance}) 
should not be larger than the present time $t_0$.

We consider the inflationary scenario with a Coleman-Weinberg 
potential of Ref.~\cite{Shafi:1983bd} with the coupling
$\lambda_3$ in Eqs.~(5) and (6) of this reference much smaller
than $\lambda_2$. In this case, Eq.~(6) of this reference 
reduces to 
\begin{equation}
A=\frac{24\lambda_2^2}{64\pi^2}\,\cdot
\end{equation} 
To be more specific, we will take as an example a particular 
viable realization of this scenario which appears in the fourth 
line of Table 4 in Ref.~\cite{okada}. In this case, the 
inflationary scale is $V_0^{1/4}\simeq 1.75\times 10^{16}~
{\rm GeV}$, which implies that $H\simeq 7.25\times 10^{13}~
{\rm GeV}$. Also $A=1.43\times 10^{-14}$ corresponding to 
$\lambda_2\simeq 6.14\times 10^{-7}$. The VEV of the inflaton 
$M$ in Ref.~\cite{Shafi:1983bd} or $v$ in Ref.~\cite{okada} is 
$M\simeq 29.4~m_{\rm P}\simeq 7.17\times 10^{19}~{\rm GeV}$. 
From the formula 
\begin {equation}
\lambda_0=A\ln\left(\lambda_2 \frac{M^2}{H^2}\right)
\end{equation} 
of Ref.~\cite{Shafi:1983bd}, we then obtain that 
$\lambda_0=1.9\times 10^{-13}$, and from Eq.~(12) of the same 
reference that 
\begin{equation}
\tau\sim\frac{\pi^2}{(8\lambda_0)^{1/2}}H^{-1}\sim 1.1\times 
10^{-7}~{\rm GeV}^{-1}.
\end{equation} 
The requirement that the defects eventually enter the horizon
(i.e. they are not inflated away) gives $\eta\lesssim 67.7$. 
From the formula $\eta=3c/\lambda_0$ of Ref.~\cite{extended}, we 
conclude that the parameter $c\sim (M_{\rm d}/M)^2\lesssim 4.3
\times 10^{-12}$, where $M_{\rm d}$ is the breaking scale 
corresponding to the defects. This scale should then satisfy 
the inequality $M_{\rm d}\lesssim 1.5\times 10^{14}~{\rm GeV}$. 
In the case of strings ($M_{\rm d}\equiv M_{\rm s}$), this 
gives for the dimensionless string tension $G\mu_{\rm s}\simeq 
(M_{\rm s}/m_{\rm P})^2\lesssim 3.7\times 10^{-9}$.

Note that the GUT scale is given by $M_{\rm GUT}\sim \lambda_2^
{1/2}M$ (see Ref.~\cite{Shafi:1983bd}). For the particular 
example we are discussing, $M_{\rm GUT}\sim 5.6\times 10^{16}~
{\rm GeV}$. Of course, this is just an order of magnitude
estimate since we do not know the precise values of the
couplings $a$ and $b$ in the potential in Eq.~(5) of 
Ref.~\cite{Shafi:1983bd} and the GUT gauge coupling 
constant.

For models predicting the existence of topologically stable 
$Z_2$ strings, we can employ Fig.~1 of Ref.~\cite{olum}, 
which holds for strings surviving until the present time.
We see that strings with $G\mu_{\rm s}\lesssim 1.5
\times 10^{-11}$, namely $M_{\rm s}\lesssim 9.45\times 
10^{12}~{\rm GeV}$, are allowed by the current experimental 
bounds. It is important to note that topologically stable 
strings with $G\mu_{\rm s}\gtrsim 10^{-20}$ will be 
possibly measurable by LISA and BBO in the future.

The number density $n_{\rm m}$ of topologically stable 
magnetic monopoles can be estimated as in Ref.~\cite{extended}.
At production it is expected to be $\sim H^{3}$. During 
inflation, the monopoles are diluted by a factor $\exp{(-3
\eta_{\rm m})}$, where $\eta_{\rm m}$ is the number of 
$e$-foldings from the time of monopole production until the 
end of inflation. During inflaton oscillations, $n_{\rm m}$ 
is multiplied by another factor $(t_{\rm r}/\tau)^{-2}$ (this 
was not taken into account in Ref.~\cite{extended}). The 
final relative monopole number density is
\begin{equation}
r\equiv \frac{n_{\rm m}}{T_{\rm r}^3}\sim \left(\frac{H}
{T_{\rm r}}\right)^3 e^{-3\eta_{\rm m}}\left(\frac{t_{\rm r}}
{\tau}\right)^{-2}\,\cdot
\end{equation} 
Requiring that $r$ does not exceed $10^{-30}$ (the Parker
bound) \cite{parker} and for the numerical example discussed 
here, we find that $\eta_{\rm m}\gtrsim 23.5$. This implies 
that, for topologically stable monopoles, the parameter 
$c_{\rm m}\sim (M_{\rm m}/M)^2\gtrsim 1.5\times 10^{-12}$, 
and the corresponding symmetry breaking scale $M_{\rm m}
\gtrsim 8.77\times 10^{13}~{\rm GeV}$ .

Next let us consider $SO(10)$ broken via $SU(4)_c\times 
SU(2)_L\times SU(2)_R$ without topologically stable $Z_2$ 
strings. In this case, during the breaking of $SU(4)_c\times 
SU(2)_L\times SU(2)_R$ to $SU(3)_c\times U(1)_{B-L}\times 
SU(2)_L \times U(1)_R$, we have the formation of $SU(4)_c$ 
(red) and $SU(2)_R$ (blue) monopoles at a scale $M_{\rm m}$. 
These monopoles are subsequently partially diluted by 
inflation. At a breaking scale $M_{\rm s}$, where $SU(3)_c
\times U(1)_{B-L}\times SU(2)_L \times U(1)_R$ reduces to 
the SM gauge group, these monopoles are connected 
by strings forming random walks with step about the horizon size 
at subsequent times. Later, the monopoles enter the 
horizon connected in pairs by one string segment. After this 
time, the monopole pairs with the string segment behave like
pressureless matter. The strings eventually decay to gravity 
waves and the monopoles merge to form either Schwinger 
monopoles or simply annihilate if they are a red or blue 
monopole with the corresponding antimonopole.

For the analysis of this case, we follow Ref.~\cite{pana}. The
present abundance of these gravity waves is given by combining 
Eqs.~(63) and (64) of this reference:
\begin{equation}
\Omega_{\rm gw} h^2 (t_0)\sim 2
\left(\frac{2}{\Gamma}\right)^{\frac{1}{2}}
(G\mu_{\rm s})^{\frac{1}{2}}\left(\frac{3.9}{10.75}
\right)^\frac{4}{3}\left(\frac{\rho_\gamma(t_0)}
{\rho_{\rm c}(t_0)}\right)h_0^2~,
\end{equation}
where $\Gamma\sim 50$, $\rho_\gamma(t_0)$ and $\rho_{\rm c}(t_0)$ 
are the present photon and critical energy densities of the 
universe respectively, and $h_0\simeq 0.7$ is the present value 
of the Hubble parameter in units of $100~{\rm km~sec^{-1}~Mpc^{-1}}$.

As an example we take $G\mu_{\rm s}\simeq 6.7\times 10^{-14}$, 
which corresponds to $M_{\rm s}\simeq 6.3\times 10^{11}~
{\rm GeV}$. The present abundance of gravity waves, in this case, 
is $\Omega_{\rm gw} h^2\simeq 10^{-12}$. The frequency $f$ of 
these waves is given by -- see Ref.~\cite{pana} -- 
\begin{equation}
f(t_0)\sim t_{\rm H}^{-1}\left(\frac{t_{\rm d}}{t_{\rm eq}}\right)^
{\frac{1}{2}}\left(\frac{t_{\rm eq}}{t_0}\right)^{\frac{2}{3}},
\end{equation}
where $t_{\rm H}$ is the time at which the monopoles enter the 
horizon, 
\begin{equation}
t_{\rm d}\sim (\Gamma G\mu_{\rm s})^{-1}2t_{\rm H}
\end{equation}
is the decay time of the strings, and $t_{\rm eq}$ is the 
equidensity time at which matter starts dominating the universe. 
For the example under discussion, we find that $f\simeq 
10^{-4}~{\rm Hz}$ provided that $t_{\rm H}\simeq 2.27~{\rm sec}$. 
The decay time of the string segments is then $t_{\rm d}\simeq 
1.35\times 10^{12}~{\rm sec}\simeq 4.28\times 10^4~{\rm yrs}$,
which is prior to matter domination at $t_{\rm eq}=4.7\times 
10^4~{\rm yrs}$. Consequently, the above calculation, which 
requires this -- see e.g. Eq.~(65) in Ref.~\cite{pana} -- is 
consistent. 

Now the question arises under what circumstances the required 
$t_{\rm H}$ can be obtained. This will be decided by the 
monopole production and evolution. As explained above,
the mean intermonopole distance at temperature $T$ after 
reheating is given by Eq.~(\ref{distance}) with $\eta=
\eta_{\rm m}$. At horizon re-entrance of the monopoles, this 
distance should be equal to $t_{\rm H}$. Of course, $T$ in 
Eq.~(\ref{distance}) should be replaced by $T_{\rm H}$ 
corresponding to $t_{\rm H}$. For $t_{\rm H}\simeq 2.27~
{\rm sec}$ and for the SM spectrum, we 
obtain $T_{\rm H}\simeq 1.74\times 10^{-4}~{\rm GeV}$, which 
is consistently lower than the reheat temperature. Our 
requirement then gives for the monopoles $\eta_{\rm m}\simeq 
48.35$, $c_{\rm m}\simeq 3.07\times 10^{-12}$, and $M_{\rm m}
\simeq 1.26\times 10^{\rm 14}~{\rm GeV}$.

Summarizing, we see that if the breaking scale of 422 to 
$SU(3)_c\times U(1)_{B-L}\times SU(2)_L \times U(1)_R$ is 
about $1.26\times 10^{14}~{\rm GeV}$ and the subsequent 
breaking scale of this group to the SM group 
is $M_{\rm s}\simeq 6.32\times 10^{11}~{\rm GeV}$, the 
monopoles re-enter the horizon after reheating at $t_{\rm H}
\simeq 2.3~{\rm sec}$ connected by a string segment with 
$G\mu_{\rm s}\simeq 6.71\times 10^{-14}$. After $t_{\rm H}$, 
the monopole string structures behave like particles and 
the strings at $t_{\rm d}$ emit gravity waves which at 
present have frequency $f=10^{-4}~{\rm Hz}$ and 
$\Omega_{\rm gw} h^2=10^{-12}$. From Fig.~1 in 
Ref.~\cite{olum}, we see that such gravity waves will be 
perfectly detectable by LISA. The spectrum of these waves is 
expected to strongly peak around $10^{-4}~{\rm Hz}$. 

\section{Conclusions}
\label{sec:conclusion}

Grand Unified Theories with a unified (single) gauge coupling 
constant such as $SU(5)$, $SO(10)$, and $E_6$ all predict the 
existence of a topologically stable  magnetic monopole that 
carries a single unit of Dirac magnetic charge (quantized with 
respect to the electron charge). This superheavy GUT scale 
magnetic monopole also carries color magnetic charge, and this 
conclusion holds independent of the symmetry breaking pattern 
of the underlying GUT model. In $SU(5)$, this magnetic monopole 
happens to be the lightest one with mass $\sim M_{\rm GUT}/
\alpha_{\rm GUT}\sim 10^{17}~{\rm GeV}$, where $\alpha_{\rm GUT}$ 
($\sim 1/10$) is the GUT fine structure constant.

In models such as $SO(10)$ or $E_6$, where the symmetry 
breaking proceeds via one or more intermediate steps, magnetic 
monopoles can appear that carry two or three units of the Dirac 
magnetic charge and their masses are related to the intermediate 
scale. Hence they are lighter than the GUT magnetic monopole with 
one unit of charge. We have observed that intermediate mass 
($\sim 10^{14}~{\rm GeV}$ or so) magnetic monopoles and 
cosmic strings of similar mass scale may be present in our galaxy 
at an observable level. We have depicted scenarios which give 
rise to superconducting cosmic strings as well as composite 
objects including a novel type of necklace that can survive 
inflation. The gravity waves emitted by some of these topological 
defects may be observable with the space based observatory LISA.
 
\vspace{.5cm}

\noindent 
{\bf Acknowledgments.}\,
{Q.S. is supported in part by the DOE Grant 
DE-SC0013880. We thank Dr. Digesh Raut for his help with the 
figures.}

\newpage 
\def\ijmp#1#2#3{{Int. Jour. Mod. Phys.}
{\bf #1},~#3~(#2)}
\def\plb#1#2#3{{Phys. Lett. B }{\bf #1},~#3~(#2)}
\def\zpc#1#2#3{{Z. Phys. C }{\bf #1},~#3~(#2)}
\def\prl#1#2#3{{Phys. Rev. Lett.}
{\bf #1},~#3~(#2)}
\def\rmp#1#2#3{{Rev. Mod. Phys.}
{\bf #1},~#3~(#2)}
\def\prep#1#2#3{{Phys. Rep. }{\bf #1},~#3~(#2)}
\def\prd#1#2#3{{Phys. Rev. D }{\bf #1},~#3~(#2)}
\def\npb#1#2#3{{Nucl. Phys. }{\bf B#1},~#3~(#2)}
\def\np#1#2#3{{Nucl. Phys. B }{\bf #1},~#3~(#2)}
\def\npps#1#2#3{{Nucl. Phys. B (Proc. Sup.)}
{\bf #1},~#3~(#2)}
\def\mpl#1#2#3{{Mod. Phys. Lett.}
{\bf #1},~#3~(#2)}
\def\arnps#1#2#3{{Annu. Rev. Nucl. Part. Sci.}
{\bf #1},~#3~(#2)}
\def\sjnp#1#2#3{{Sov. J. Nucl. Phys.}
{\bf #1},~#3~(#2)}
\def\jetp#1#2#3{{JETP Lett. }{\bf #1},~#3~(#2)}
\def\app#1#2#3{{Acta Phys. Polon.}
{\bf #1},~#3~(#2)}
\def\rnc#1#2#3{{Riv. Nuovo Cim.}
{\bf #1},~#3~(#2)}
\def\ap#1#2#3{{Ann. Phys. }{\bf #1},~#3~(#2)}
\def\ptp#1#2#3{{Prog. Theor. Phys.}
{\bf #1},~#3~(#2)}
\def\apjl#1#2#3{{Astrophys. J. Lett.}
{\bf #1},~#3~(#2)}
\def\apjs#1#2#3{{Astrophys. J. Suppl.}
{\bf #1},~#3~(#2)}
\def\n#1#2#3{{Nature }{\bf #1},~#3~(#2)}
\def\apj#1#2#3{{Astrophys. J.}
{\bf #1},~#3~(#2)}
\def\anj#1#2#3{{Astron. J. }{\bf #1},~#3~(#2)}
\def\mnras#1#2#3{{MNRAS }{\bf #1},~#3~(#2)}
\def\grg#1#2#3{{Gen. Rel. Grav.}
{\bf #1},~#3~(#2)}
\def\s#1#2#3{{Science }{\bf #1},~#3~(#2)}
\def\baas#1#2#3{{Bull. Am. Astron. Soc.}
{\bf #1},~#3~(#2)}
\def\ibid#1#2#3{{\it ibid. }{\bf #1},~#3~(#2)}
\def\cpc#1#2#3{{Comput. Phys. Commun.}
{\bf #1},~#3~(#2)}
\def\astp#1#2#3{{Astropart. Phys.}
{\bf #1},~#3~(#2)}
\def\epjc#1#2#3{{Eur. Phys. J. C}
{\bf #1},~#3~(#2)}
\def\nima#1#2#3{{Nucl. Instrum. Meth. A}
{\bf #1},~#3~(#2)}
\def\jhep#1#2#3{{J. High Energy Phys.}
{\bf #1},~#3~(#2)}
\def\jcap#1#2#3{{J. Cosmol. Astropart. Phys.}
{\bf #1},~#3~(#2)}
\def\lnp#1#2#3{{Lect. Notes Phys.}
{\bf #1},~#3~(#2)}
\def\jpcs#1#2#3{{J. Phys. Conf. Ser.}
{\bf #1},~#3~(#2)}
\def\aap#1#2#3{{Astron. Astrophys.}
{\bf #1},~#3~(#2)}
\def\mpla#1#2#3{{Mod. Phys. Lett. A}
{\bf #1},~#3~(#2)}

\end{document}